\documentclass[twocolumn, prl]{revtex4}
\usepackage{graphicx}
\usepackage{latexsym}

\newcommand{\be}{\begin{equation}}
\newcommand{\ee}{\end{equation}}
\newcommand{\bea}{\begin{eqnarray}}
\newcommand{\eea}{\end{eqnarray}}

\begin{document}
\title{Hopping conductivity of a suspension of nanowires in an insulator}
\author{Tao Hu, B. I. Shklovskii}
\affiliation{Department of Physics, University of Minnesota \\
116 Church Street SE, Minneapolis, MN 55455}
\date{\today}

\begin{abstract}

We study the hopping conduction in a composite made of straight
metallic nanowires randomly and isotropically suspended in an
insulator. Uncontrolled donors and acceptors in the insulator lead
to random charging of wires and, hence, to a finite bare density of
states at the Fermi level. Then the Coulomb interactions between
electrons of distant wires result in the soft Coulomb gap. At low
temperatures the conductivity is due to variable range hopping of
electrons between wires and obeys the Efros-Shklovskii (ES) law
$\ln\sigma \propto -(T_{ES}/T)^{1/2}$. We show that $T_{ES} \propto
1/(nL^3)^2$, where $n$ is the concentration of wires and $L$ is the
wire length. Due to enhanced screening of Coulomb potentials, at
large enough $nL^3$ the ES law is replaced by the Mott law.

\end{abstract}
\maketitle

In recent years, electron transport in composites made of metallic
granules surrounded by some kind of insulator generated a lot of
interest in both fundamental and applied research. Conductivities of
the composites with spherical or generally speaking single scale
granules are well studied (see review~\cite{Vinokur} and references
therein). If the number of granules $n$ per unit volume is large,
granules touch each other and the conductivity is metallic. When $n$
is smaller than the percolation threshold $n_c$, granules are
isolated from each other, the composite is on the insulating side of
metal-insulator transition and the conductivity is due to hopping.
It was observed that in this case, the temperature dependence of the
conductivity $\sigma$ obeys
\be \sigma = \sigma_0\exp[-(T_0/T)^{\alpha}], \ee
where $T_0$ is a characteristic temperature and $\alpha = 1/2$. This
behavior of conductivity $\sigma$ was interpreted as the
Efros-Shklovskii (ES) variable range hopping (VRH) between
granules~\cite{Zhang,Vinokur}. The finite bare density of states at
the Fermi level was attributed to the charged impurities in the
insulator playing the role of randomly biased gates. Then the
interaction of electrons residing on different dots creates the
Coulomb gap, which leads to ES variable range hopping.

\begin{figure}
\includegraphics[width=0.35\textwidth]{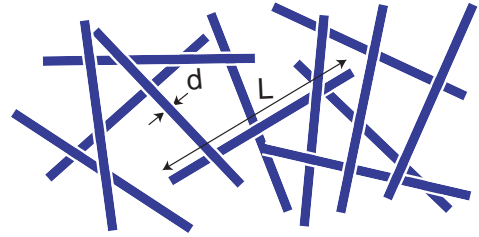}
\caption{A suspension of neutral metallic wires in an insulator with
concentration $1/L^3 \ll n \ll 1/L^2d$.} \label{fig:suspension}
\end{figure}

For needle-like granules such as metallic whiskers or metallic
nanotubes, the experimental situation is more complicated than for
single scale ones. Recently ES temperature dependence of the
conductivity was reported~\cite{Benoit} for composites made of
bundles of single-wall carbon nanotubes (SWNT) suspended in
insulating polymers. On the other hand, the Mott law with
$\alpha=1/4$ was also observed in many nanotube based
materials~\cite{Benoit,Gaal,Fuhrer,Yosida}. In order to understand
the puzzling crossover between ES and Mott hopping, in this paper,
we study the low temperature ohmic dc transport in an isotropic
suspension of elongated metallic granules in an insulating medium
(see Fig. \ref{fig:suspension}). Below we call such a granule a
\emph{wire} for brevity.

In this paper we restrict ourselves to scaling approximation for the
conductivity and to delineating the corresponding scaling regimes.
In our scaling theory, we drop away both all numerical factors and,
moreover, also all the logarithmic factors, which do exist in this
problem, because it deals with strongly elongated cylinders. In this
context, we will use symbol ``$=$'' to mean ``equal up to a
numerical coefficient or a logarithmic factor''.

\begin{figure}
\includegraphics[width=0.35\textwidth]{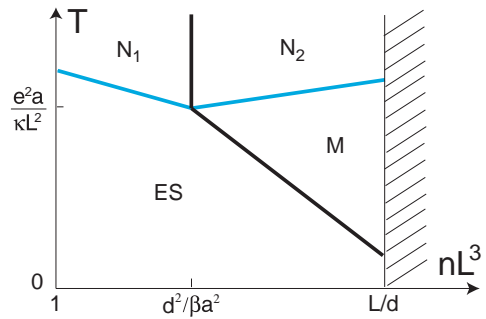}
\caption{(Color online) The summary of the transport regimes for
relatively clean wires schematically plotted in the plane of $nL^3$
and temperature $T$. The dark line separates regimes where Coulomb
interaction is important ($\rm N_1$ [Eq. (\ref{eq:A1})] and ES law
[Eq. (\ref{eq:ES})]) from regimes where it plays a minor role ($\rm
N_2$ [(\ref{eq:A2})] and Mott (M) law [Eq. \ref{eq:Mott}]). The grey
(blue) line separates activated the nearest neighbor hopping (NNH)
regimes from the VRH regimes. Shaded domain represents the metallic
side of the insulator-metal transition. Instead of smooth crossovers
at dark and grey lines, in the vicinity of this transition, the
conductivity has a critical behavior.} \label{fig:regime}
\end{figure}

We assume each wire is a cylinder with the length $L$ and the
diameter $d \ll L $. When the concentration of wires $n$ is very
small, the distance between them is much larger than $L$, we can
easily apply the results obtained in the Ref.~\cite{Zhang}. Here we
consider a range of concentrations $1/L^3 \ll n \ll 1/L^2d$, where
$n$ is so large that spheres built on each wire as the diameter
strongly overlap, but the system is still far below the percolation
threshold $n_c = 1/L^2d$. Percolation threshold at very small
concentration $n_c = 1/L^2d$ is the result of large excluding volume
$L^2d$ which one wire creates for centers of others. (At
concentrations above $1/L^2d$ it also becomes impossible to place
wires randomly and isotropically because of nematic ordering
\cite{Onsager}, but there is no such problem at $n \ll n_c$). Our
results are summarized in Fig. \ref{fig:regime} in the plane of
parameters $T$ and $nL^3$. The main result is that at low
temperatures, we arrive at ES law
\be \sigma = \sigma_0 \exp[-(T_{ES}/T)^{1/2}]\label{eq:ES}\ee
with
\be T_{ES} = \frac{e^2}{\kappa a(nL^3)^2}, \label{eq:TES} \ee
where $\kappa$ is the dielectric constant and $a$ is the tunneling
decay length in the insulator. The large factor $(nL^3)^2$ in Eq.
(\ref{eq:TES}) is a result of the enhancement of both the effective
dielectric constant and the localization length in the composite. At
large enough $nL^3$ and higher temperatures, ES law is replaced by
the Mott law. And when the temperatures are sufficiently high, both
VRH regimes are replaced by the activated nearest neighboring
hopping (NNH) regimes.

Let us start from the dielectric constant of the composite. In the
range of concentrations $1/L^3 \ll n \ll 1/L^2d$, its macroscopic
dielectric constant $\kappa_{\rm eff}$ is greatly enhanced due to
polarization of long metallic wires. It was shown in the
Ref~\cite{Sarychev,conductivity}, that
\be \kappa_{\rm eff} = (nL^3)\kappa. \label{eq:kappa} \ee
Such result can be understood in the following way. If the wave
vector $q$ is larger than $1/L$, the static dielectric function for
the wire suspension has the metal-like form
\be \epsilon(q) = \kappa\left(1+\frac{1}{q^2r_{s}^{2}}\right), \ee
where $r_{s}= (nL)^{-1/2}$ is the typical separation of the wire
from other wires. It plays the role of screening radius for the wire
charge. The function $\epsilon(q)$ grows with decreasing $q$ until
$q = 1/L$ where the composite loses its metallic response and
$\epsilon(q)$ saturates. As a result, the macroscopic effective
dielectric constant is given by $\kappa_{\rm eff} = \epsilon(q =
1/L) = nL^3\kappa$. Within our scaling approximation, one can also
derive Eq. (\ref{eq:kappa}) starting from the facts that each
isolated wire has polarizability equal to $L^3$ and due to random
positions of wires the acting electric field does not differ from
the average one.

The spacing between charging energy levels of such wire is the order
of $E_c = e^2/\kappa_{\rm eff}L = e^2/\kappa nL^4$. If the wire is
thick enough, it can be treated as a three dimensional object and
the mean quantum level spacing $\delta$ in the wire is the order of
$\lambda_F\hbar^2/mLd^2$, where $\lambda_F$ is the Fermi wavelength
and $m$ is the effective electron mass. Narrow wires such as SWNT
with single conducting channel will be discussed later. Using the
tunneling decay length $a$ in the insulator, we can rewrite $\delta$
in the form $\delta = \beta e^2 a^2/\kappa d^2L$ where $\beta =
\lambda_F/a$. The ratio $E_c/\delta$ decreases proportional to
$1/nL^3$ because of enhanced screening of Coulomb interaction in the
denser system of longer wires.

\begin{figure}
\includegraphics[width=0.45\textwidth]{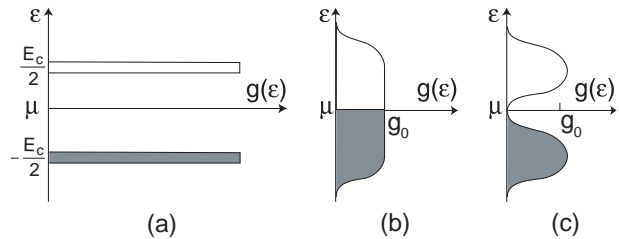}
\caption{The density of ground states at $E_c \gg \delta$. (a) BDOGS
in neutral wires, (b) BDOGS  for charged wires with large enough
concentration of donors in the insulator in the absence of the long
range Coulomb interaction; (c) The density of ground states of
charged wires with the Coulomb gap in the vicinity of the Fermi
level $\mu$. Occupied states are shaded.} \label{fig:BDOGS}
\end{figure}

We start our discussion with relatively small $nL^3$ such that
$nL^3<d^2/\beta a^2$ and $E_c \gg \delta$. The density of states and
localization length are all the information we need to calculate the
VRH conductivity. In contrary to a doped semiconductor, in a
suspension of neutral wires, the bare density of states at the Fermi
level $g_0=0$. Charging energy levels of the suspended wires are
shown in Fig. \ref{fig:BDOGS}(a), they are equally spaced by $E_c$.
The Fermi level is at zero energy between empty and filled levels.

Since the ground states of wires determine the low temperature
hopping transport, we consider only the ground state in each wire at
a given number of electrons and exclude excited states~\cite{Zhang}.
Thus, the density of states we need can be called bare density of
ground states (BDOGS). According to Ref.~\cite{Zhang}, the finite
BDOGS near the Fermi level originates from uncontrollable donors (or
acceptors) in the insulating host. Donors have the electron energy
above the Fermi energy of wires. Therefore, they donate electrons to
wires. A positively charged ionized donor can attract and
effectively bind fractional negative charges on all neighboring
wires, leaving the rest of each wire fractionally charged. At a
large enough average number ($\gg 1$) of donors per wire, effective
fractional charges on different wires are uniformly distributed from
$-e/2$ to $e/2$. In such a way the Coulomb blockade in a single wire
is lifted and the discrete BDOGS get smeared (see Fig.
\ref{fig:BDOGS}(b)). The BDOGS $g_0$ becomes $1/E_cL^3$. In the very
vicinity of the Fermi energy, the long range Coulomb interaction
creates the parabolic Coulomb gap $\Delta$ (see Fig.
\ref{fig:BDOGS}(c)). The Coulomb gap affects energy interval
$|\epsilon| \lesssim \Delta$. At low enough $T$, when the range of
energy levels around the Fermi level responsible for hopping
$(T_{ES}T)^{1/2}$ is much smaller than $\Delta$, the conductivity
obeys the ES law (see Eq. (\ref{eq:ES})) and does not depend on
$g_0$~\cite{ES}. The parameter $T_{ES} = e^2/\kappa_{\rm eff} \xi$,
where $\xi$ is the localization length for tunneling to distances
much larger than $L$. Let us concentrate now on the nontrivial value
of the localization length $\xi$.

\begin{figure}
\includegraphics[width=0.45\textwidth]{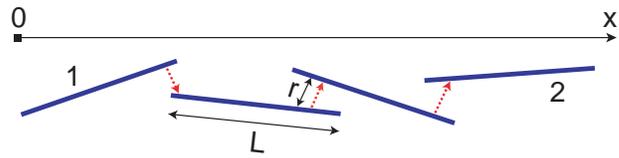}
\caption{(Color online) An illustration of a long range ($\gg L$)
hop from the wire 1 to the wire 2 via a sequence of wires.}
\label{fig:tunneling}
\end{figure}

It was suggested~\cite{Boris} that such long range hopping process
can be realized by tunneling through a sequence of wires (see Fig.
\ref{fig:tunneling}). The states of the intermediate wires
participate in the tunneling process as \emph{virtual states}.
Nowadays the virtual electron tunneling through a single granule is
called
co-tunneling~\cite{Nazarov,Ioselevich,Beloborodov,Vinokur,Fogler1}
and regarded as a key mechanism of low temperature charge transport
via quantum dots. One should distinguish the two co-tunneling
mechanisms, elastic and inelastic. During the process of elastic
co-tunneling, the electron tunneling through an intermediate virtual
state in the granule leaves the granule with the same energy as its
initial state. On the contrary, the tunneling electron in the
inelastic co-tunneling mechanism leaves the granule with an excited
electron-hole pair behind it. Which mechanism dominates the
transport depends on the temperature. Inelastic co-tunneling
dominates at $T>\sqrt{E_c\delta}$, while below this temperature,
elastic co-tunneling wins. In discussions below, we consider
sufficiently low temperatures such that only elastic co-tunneling is
involved.

First we remind the idea of calculation of the localization length
$\xi$ used in Ref.~\cite{Boris}. (Such approach was also applied in
the papers~\cite{Zhang,Fogler}). When an electron tunnels through
the insulator between wires at the nearest neighboring hopping (NNH)
distance $r$, it accumulates dimensionless action $r/a$, where $a$
is the tunneling decay length in the insulator. We want to emphasize
that the NNH distance $r$ is realized only in one point of the wire
and, therefore, is much shorter than the typical separation along
the wire from other wires $r_s$. NNH distance $r$ can be calculated
using the percolation method~\cite{ES}. If one builds around each
wire a cylinder with length $L$ and radius $r$, percolation through
these cylinders appears, when $nL^2r = 1$. This happens because
excluded volume created by one cylinder is not $r^2L$, but $L^2r$.
As a result, we obtain $r = 1/nL^2$. We assume the metallic wire is
only weakly disordered and hence we neglect the decay of electron
wave function in the wire. (This assumption will be relaxed later.)
Over distance $x$, electron accumulates $x/L$ additive actions of
the order of $r/a$. Thus, its wave function decays as
$\exp(-xr/La)=\exp(-x/\xi)$, where the localization length
$\xi=aL/r=a(nL^3)$. This contributes another factor $nL^3$ to Eq.
(\ref{eq:TES}). Eq. (\ref{eq:ES}) with $T_{ES}$ given by Eq.
(\ref{eq:TES}) is valid in the left lower corner of Fig.
\ref{fig:regime} labeled as ES. More rigorous
calculation~\cite{Beloborodov} of the localization length $\xi$
gives $\xi = L/[\ln(E_c\pi/\delta)+1/nL^2a]$. Our simple derivation
of $\xi$ along the line of Ref.~\cite{Boris} has lost the first term
in the denominator. We argue that in our case $1/nL^2a \gg d/a$, and
$d/a$ can be quite large. As a result, within a large range of the
ratio $E_c/\delta$, the term $\ln(E_c\pi\delta)$ is only a small
correction to the leading term $1/nL^2a$.

At high temperatures, the conductivity is dominated by the nearest
neighbor hopping. Using the NNH distance $r$, we obtain the
conductivity with activated $T$-dependence,
\be \sigma \propto {\rm e}^{-1/nL^2a}{\rm e}^{-E_A/T},
\label{eq:A}\ee
where the activation energy $E_A$ is determined by the charging
energy
\be E_A = \frac{e^2}{\kappa_{\rm eff}L} = \frac{e^2}{\kappa nL^4} \
\ \ {\rm if} \ \ \ \delta \ll E_c. \label{eq:A1} \ee
The range of validity of Eqs. (\ref{eq:A}) and (\ref{eq:A1}) is
shown in the upper left corner of Fig. \ref{fig:regime} and labeled
as $\rm N_1$.

When $L$ becomes so large that $\delta > E_c$, the BDOGS $g_0$
evolves from $(E_cL^3)^{-1}$ to $g_0 = (\delta L^3)^{-1} = \kappa
d^2/e^2\beta L^2a^2$. Since $g_0$ decreases and $\kappa_{\rm eff}$
grows with $L$, the width of the Coulomb gap $\Delta$, which can be
estimated as $= \sqrt{g_0e^6/\kappa_{\rm eff}^3}$ decreases with
$L$. Eventually, the Coulomb gap becomes narrower than the width of
Mott's optimal band $\epsilon_M = T^{3/4}/(g_0\xi^3)^{1/4}$ at given
$T$. At this point, the ES law is replaced by the conventional
Mott's law,
\be \sigma = \sigma_0 \exp[-(T_M/T)^{1/4}], \label{eq:Mott} \ee
where
\be T_M = \frac{1}{g_0\xi^3} = \frac{\beta e^2}{\kappa a n^3L^7d^2}.
\label{eq:TM} \ee
This regime is labeled as M in Fig. \ref{fig:regime}. It crossover
to the ES regime at $T = e^2d^2/\kappa a \beta nL^5$, which can be
obtained by equating $\Delta$ and $\epsilon_M$. This change of the
temperature dependence of the conductivity from $1/2$ ES law to
$1/4$ Mott law was observed at relatively high temperatures and
large concentrations in the experiment~\cite{Benoit}.

At high temperatures, the Mott's law is replaced by the NNH
conductivity with activation energy equal to the spacing of quantum
levels $\delta$. The $T$-dependence of the conductivity is,
\be \sigma \propto
\exp\left(-\frac{1}{nL^2a}\right)\exp\left(-\frac{e^2\beta
a^2}{\kappa d^2LT}\right), \ \ \ {\rm if} \ \ \ \delta \gg E_c.
\label{eq:A2} \ee
We call this regime $\rm N_2$ and show it in the right upper corner
of Fig. \ref{fig:regime}. It is easy to check that ES and Mott VRH
transport regimes match corresponding NNH regimes $\rm N_1$ and $\rm
N_2$ smoothly, when VRH distance is of the order of the wire length
$L$. This happens at the gray line of Fig. \ref{fig:regime}. The
Coulomb regimes ES and $\rm N_1$ smoothly crossover to the regimes M
and $\rm N_2$ of noninteracting electrons at the dark line of Fig.
\ref{fig:regime}. We stop at $nL^3 = L/d$, where percolation leads
to the insulator-metal transition. Our scaling approach is not
designed to study the critical behavior of the conductivity in the
vicinity of the transition.

\begin{figure}
\includegraphics[width=0.4\textwidth]{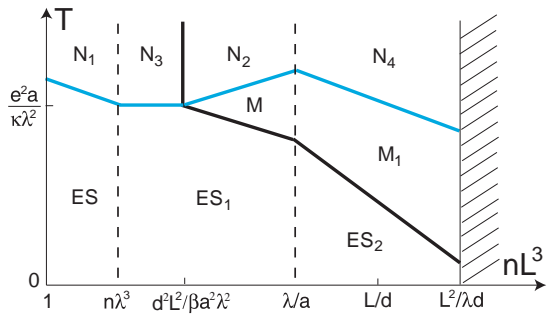}
\caption{(Color online) Transport regimes for wires, where electrons
are localized at length $\lambda < L$. Dark line separates regimes
where Coulomb interaction is important ($\rm N_{1,3}$ and ES laws)
from regimes where it plays minor role ($\rm N_{2,4}$ and Mott
laws). The grey (blue) line separates activated NNH regimes from VRH
regimes. The left dashed line marks the border where dielectric
constant $\kappa_{\rm eff}$ starts growing with $nL^3$, while the
right dashed line serves as the border where localization length
$\xi$ stops growing and becomes $\lambda$. The shaded domain
represents the metallic side of the insulator-metal transition.
Instead of smooth crossover, in the vicinity of the transition, the
conductivity has a critical behavior.} \label{fig:disorder}
\end{figure}

Now we switch to the case of metallic wires with a strong internal
disorder, where the electron wave function decay length $\lambda$ is
much smaller than the wire length $L$. The summary of transport
regimes for this case is presented in Fig. \ref{fig:disorder}. In
order to show all the possible regimes, we assume that $\lambda$ is
small enough for the order of labels on the $nL^3$ axis shown in
Fig. \ref{fig:disorder}. (At a larger $\lambda$ only some of the
shown regimes may exist.) We also plotted regimes labeled as ES and
$\rm N_1$ on the left of the border line $nL^3 = n\lambda^3$. The
characteristic energies of electron transport in these regimes are
given by Eqs. (\ref{eq:TES}) and (\ref{eq:A1}) respectively.
When $L\gg \lambda$, the wires can be treated as a larger number of
clean metallic wires with length of the order $\lambda$ and the
concentration $nL/\lambda$. Following the logic of
Ref.~\cite{Sarychev} and our derivation of $\kappa_{\rm eff}$ (Eq.
(\ref{eq:kappa})) one can show that the effective macroscopic
dielectric constant is given by
\be \kappa_{\rm eff1} = \kappa n L\lambda^2. \label{eq:kappa1} \ee
Comparing to (Eq. (\ref{eq:kappa})) we see that the dielectric
constant $\kappa_{\rm eff1}$ is strongly reduced by the internal
disorder of wires. This happens because the disorder effectively
"cuts" the wire into short pieces albeit increases the concentration
of the short wires.

The internal disorder also changes the localization length for the
long distance ($\gg L$) tunneling. Using the argument we made above,
we obtain
\be \frac{1}{\xi} = \frac{1}{\lambda} + \frac{1}{nL^3a}, \ee
where the first term on the right hand side accounts for the decay
of the wave function in the wire, while the second term represents
the decay in the insulating medium between the wires. When $nL^3 <
\lambda/a$, the second term dominates, $\xi = nL^3a$, which is the
same as what we got for the case of clean wire. However, because of
new dielectric constant $\kappa_{\rm eff1}$ (Eq. \ref{eq:kappa1}),
we obtain new regime labeled $\rm ES_1$ in the Fig.
\ref{fig:disorder}. Here the conductivity obeys ES law with the
characteristic temperature
\be T_{ES_1} = \frac{e^2}{\kappa a n^2L^4\lambda^2}. \ee
The charging energy $E_c$ also changes to $e^2/\kappa_{\rm eff1}L =
e^2/\kappa nL^2\lambda^2$. It decreases with growing $nL^3$ and
reaches the mean quantum level spacing $\delta$ of the wire at $nL^3
= d^2L^2/\beta a^2\lambda^2$. With relatively high temperature and
small $nL^3<d^2L^2/\beta a^2\lambda^2$, the regime $\rm ES_1$
crossover to the regime $\rm N_3$, where the conductivity can be
represented by the same equation (\ref{eq:A}) but with a different
activation energy $E_A = e^2/\kappa nL^2\lambda^2$. When
$nL^3>d^2L^2/\beta a^2\lambda^2$, the Coulomb interaction plays a
minor role. Hopping conductivities in regimes $\rm M$ and $\rm N_2$
are not affected by the change of the dielectric constant and are
given by the same Eqs. (\ref{eq:Mott}), (\ref{eq:TM}) and
(\ref{eq:A}), (\ref{eq:A2}) respectively.

At even larger $nL^3 > \lambda/a$, the decay of electron wave
functions in the wire dominates, thus $\xi = \lambda$. As a result,
we have another ES regime, $\rm ES_2$, with
\be T_{ES_2} = \frac{e^2}{\kappa nL\lambda^3}, \ee
the Mott regime $\rm M_1$ with
\be T_{M_1} = \frac{e^2}{\kappa a n^2L^4\lambda^2}. \ee
and the activated nearest neighboring hopping regime $\rm N_4$, with
the activation energy $\delta$ and
\be \sigma \propto
\exp\left(-\frac{L}{\lambda}\right)\exp\left(-\frac{e^2\beta
a^2}{\kappa d^2LT}\right). \ee

Thus, we have completed our consideration of the phase diagram up to
$nL^3 = L/d$, at which wires begin to percolate. What happens to the
conductivity if one manages to create an isotropic suspension with
$nL^3 > L/d$? In contrary to relatively clean wires, there is no
insulator-metal transition around $nL^3 = L/d$, where wires start to
touch each other. Indeed such transition happens when the typical
branch of the percolating network is metallic, which means the
typical distance between two nearest neighboring contacts along the
same wire can not be $L$, but should be smaller than $\lambda$. We
can now think about percolation over pieces of wires with the length
$\lambda$ and the effective concentration $nL/\lambda$. Such
percolation appears at $(nL/\lambda)(\lambda^2d)=1$ or $nL^3 =
L^2/\lambda d$. This is the threshold of insulator-metal transition.
As a result our phase diagram of hopping regimes is extended as far
as $nL^3 = L^2/\lambda d$.

In this paper we concentrated on relatively thick wires with the
three dimensional density of states. Narrow wires with as few as one
conducting channel should be treated as one dimensional objects. In
such case the level spacing is given by $\delta =
\hbar^2/m\lambda_FL$ and thus $E_c/\delta = (1/nL^3)(\lambda_F/a) =
\beta/nL^3$. Since most probably $\beta$ is the order of $1$, while
$nL^3 \gg 1$, the charging energy $E_c$ is always smaller than level
spacing $\delta$. Coulomb interaction plays minor role in such
system. Therefore the left part of our phase diagram Fig.
\ref{fig:regime} including ES regime vanishes.

It is tempting to apply this theory to carbon nanotubes. Metallic
SWNT has only one conducting channel. Therefore for them, the domain
of ES hoping most likely vanishes. Bundles of SWNT, studied in
Ref.~\cite{Benoit} definitely have many channels and our theory
should work for them providing the geometry of bundles may be
approximated by a set of randomly oriented rigid rods of a given
characteristic length $L$. In this case, the range of ES
conductivity should exist. ES range may also exist for multi-wall
nanotubes, where one can expect many parallel metallic channels.

In this paper, we considered only straight wires. One can generalize
this theory to random system of flexible wires formed by
interpenetrating Gaussian coils (for example conducting polymers
suspended in insulating medium). This theory will be published
elsewhere.

We are grateful to M. Fogler, M. Foygel, Y.M. Galperin and A.V.
Lopatin for useful discussions.

\end{document}